# Mapping the complete reaction path of a complex photochemical reaction

Adam D. Smith,[1,†] Emily M. Warne,[1,†] Darren Bellshaw,[2] Daniel A. Horke,[3,4] Maria Tudorovskya,[2] Emma Springate,[5] Alfred J. H. Jones,[5] Cephise Cacho,[5] Richard T. Chapman,[5] Adam Kirrander,[2]* and Russell S. Minns[1]*

[1]Chemistry, University of Southampton, Highfield, Southampton SO17 1BJ, UK

[2]EaStCHEM, School of Chemistry, University of Edinburgh, David Brewster Road, Edinburgh EH9 3FJ, United Kingdom.

[3]Center for Free-Electron Laser Science, Deutsches Elektronen-Synchrotron DESY, Notkestrasse 85, 22607 Hamburg, Germany

[4]The Hamburg Centre for Ultrafast Imaging, Universität Hamburg, Luruper Chaussee 149, 22761 Hamburg, Germany

[5]Central Laser Facility, STFC Rutherford Appleton Laboratory, Didcot, Oxfordshire OX11 0QX, UK

*Correspondence to:  adam.kirrander@ed.ac.uk and r.s.minns@soton.ac.uk

† These authors contributed equally to this work.

**Abstract**: We probe the dynamics of dissociating $CS_2$ molecules across the entire reaction pathway upon excitation. Photoelectron spectroscopy measurements using laboratory-generated femtosecond extreme ultraviolet pulses monitor the competing dissociation, internal conversion, and intersystem crossing dynamics. Dissociation occurs either in the initially excited singlet manifold or, via intersystem crossing, in the triplet manifold. Both product channels are monitored and show that despite being more rapid, the singlet dissociation is the minor product and that triplet state products dominate the final yield. We explain this by consideration of accurate potential energy curves for both the singlet and triplet states. We propose that rapid internal conversion stabilises the singlet population dynamically, allowing for singlet-triplet relaxation via intersystem crossing and efficient formation of spin-forbidden dissociation products on longer timescales. The study demonstrates the importance of measuring the full reaction pathway for defining accurate reaction mechanisms.

Photochemical processes involve complex coupled motion of electrons and nuclei on fast timescales, leading to a dynamic flow of energy between nuclear and electronic degrees of freedom and, through competing processes, to the eventual formation of reaction products. The ultraviolet (UV) photochemistry of $CS_2$ is a benchmark reaction, for which a clear picture of the dynamics and a full explanation for timescales and photoproduct branching ratios remains elusive despite the apparent structural simplicity of the molecule. The origin of the complexity is the same as in larger systems: a high number of near degenerate electronic states, rapid internal conversion (IC), and intersystem crossing (ISC) dynamics. [1–8] In $CS_2$, eventual dissociation products are associated with either singlet or triplet spin states of the dissociated S atom. The dissociation is fast, on the order of a few hundred femtoseconds. Surprisingly, extensive experiments have shown that the spin-forbidden triplet state product dominates the dissociation yield over the more direct spin-allowed singlet state dissociation. [9–12] Providing a mechanistic explanation for this reaction outcome is difficult since experimental techniques with the required sensitivity to the electronic and nuclear dynamics struggle to measure the full reaction pathway, making it necessary to infer the mechanism from partial measurements that may have missed key steps.

In seminal work, Stolow and co-workers [2,3] used molecular alignment in combination with UV photoelectron imaging to measure the early-time dynamics and the correlated changes in electronic structure and molecular geometry in $CS_2$, highlighting the importance of the initial IC dynamics. More recent measurements by Suzuki and co-workers [5,6] mapped the bending vibrational motion and the large associated change in ionisation energy, and observed the appearance of the singlet dissociation products through an accidental resonance with autoionising states of the S atom produced using a 9.3 eV probe. [8] Even with such a high-energy probe, the dynamics in the dominant triplet state dissociation could not be observed and the mechanism for singlet dissociation was difficult to define. Extensive experiments also monitored the angular and velocity distributions of the products [12–14] and provided branching ratios for the triplet to singlet dissociation channels of around 3:1 (however estimates over the years have varied from 0.25:1 to 6:1 [9–12,15–18]). Importantly, the experiments have provided few explanations for why the spin forbidden product dominates the dissociation yield.

Using table-top femtosecond extreme ultraviolet (XUV) pulses from a high harmonic generation (HHG) source as a photoelectron spectroscopy (PES) probe, we measure the entire reaction pathway in sufficient detail to identify the most important electronic and structural transformations associated with the reaction. The energy of the XUV probe allows us to ionise and measure all of the relevant electronic states and molecular geometries associated with the competing dissociation reactions in $CS_2$. Being based on HHG, the experiments have the potential to combine attosecond time resolution [19–22] with PES's proven sensitivity to both electronic and geometric structure changes. [23] By monitoring all aspects of the dynamics, we identify the processes that control the branching between the singlet and triplet dissociation products and provide a mechanism for the full dissociation process, with the experimental results supported by accurate potential energy curves for the excited singlet and triplet states.

Experimental details are provided in the supplementary information (SI). [24] Briefly, A 200 nm (6.2 eV) pump pulse excites $CS_2$ from the ground state predominantly into the $3^1A'$ ($^1B_2{}^1\Sigma_u^+$) excited state (see Fig. S4 in SI for transition dipole moments). The subsequent dynamics, which involves rapid IC and ICS, is probed at a series of time-delays via ionisation by an isolated high harmonic at 57.4 nm (21.6 eV). The instrument response function given by the cross-correlation of the pump and probe is 180 fs. The resulting time-dependent photoelectron spectra provides a sequence of instantaneous snapshots of the molecule all the way to the dissociation products,

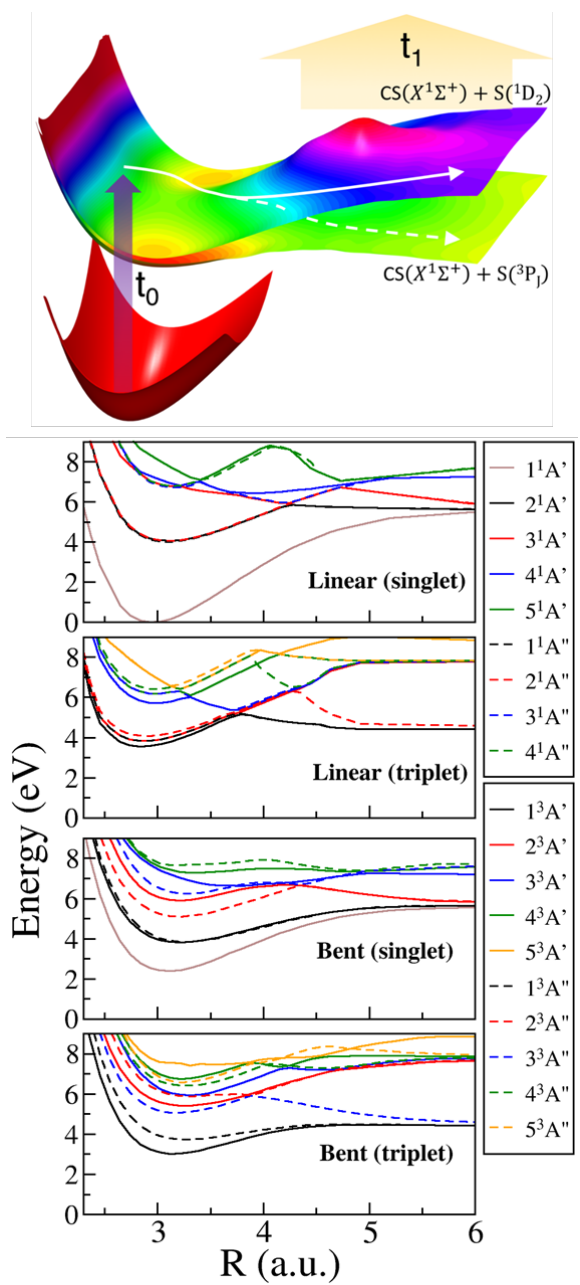

**Figure 1**. Overview of the potential energy surfaces of CS$_2$ leading to dissociation, with a 3D rendering of one singlet and one triplet potential energy surface and accurate potential energy curves of the relevant electronic states are shown at linear and bent geometry along the C–S bond stretching coordinate. The second C–S bond is kept at equilibrium length $R_{CS} = 1.569$ Å. Solid lines represent states of A′ symmetry, dashed lines represent states of A″ symmetry. From top to bottom: Linear (178°) singlets; linear (178°) triplets; bent (120°) singlets; bent (120°) triplets. Several states are (near-)degenerate at linear geometry. The potential energies are calculated at the MRCI(14,10)/aug-cc-pVTZ level (further details and angular potentials in the SI).

$$\text{CS}_2 + h\nu \text{ (200 nm)} \rightarrow \begin{cases} \text{CS}(X^1\Sigma^+) + \text{S}(^1D_2) \\ \text{CS}(X^1\Sigma^+) + \text{S}(^3P_J) \end{cases}.$$

The competition between the two product pathways derives from coupling between electronic and nuclear motion and results in bond-breaking and formation of sulphur atoms in two different spin states. The evolution of the valence-electronic structure is complex and a challenge to experiment and theory such that, despite extensive study, no consistent mechanistic picture has been derived. This is reflected in Figure 1 where we plot radial potential energy curves for the $CS_2$ molecule, with the spin-orbit coupling making it necessary to consider both the singlet *and* triplet manifolds (Fig. S3 in SI also show angular potential energy curves). The near-degeneracy of the initially excited $3^1A'$ state with the $4^1A'$ state at linear geometries means that population can be rapidly redistributed. As the molecule bends and stretches this near-degeneracy is lost and the triplet states become energetically and dynamically accessible, increasing the chance of ISC. Importantly, the barrier to dissociation in the excited singlet states is higher than those associated with the triplet dissociation process.

In Figure 2A we plot the probe-only photoelectron spectrum of ground state $CS_2$ (black line). The binding energy axis is calculated as the difference between the probe photon energy and the measured electron kinetic energy. The peaks in the spectrum correspond to ionisation of the $CS_2$ ($\tilde{X}\,^1\Sigma_g^+$) ground state into the $\tilde{X}\,^2\Pi_g$ (10.06 eV), $\tilde{A}\,^2\Pi_u$ (12.7 eV) and $\tilde{B}\,^2\Sigma_u^+$ (14,5 eV), states of the ion. [25] The ground state spectrum is taken as the background and subtracted from all spectra obtained at later times. Two background-subtracted spectra are also plotted in Figure 2A. The plotted spectra are averages obtained between 40 – 540 fs, corresponding to the early stages of the reaction, and between 4.04 – 10.54 ps, which corresponds to times when the dissociation process is complete. Negative signals correspond to a depletion of the signal at that energy and positive features to an enhancement.

The background-subtracted spectrum at long delay-times shows a strong depletion of the peaks associated with the ground state, and several new features corresponding to the dissociation products. Peaks associated with ionisation of the ground state $CS(X\,^1\Sigma^+)$ fragment and both ground ($^3P$) and excited state ($^1D$) sulphur atoms are labelled by combs in Figure 2A based on known ionisation energies, [26–28] (see SI [24]). At earlier times, we also see features at lower binding energies that correlate with the excited states of bound $CS_2$. These features are of lower intensity due to their smaller ionisation cross-section, but appear with zero background which aids their detection. The features associated with the excited states are much broader due to the strong dependence of the ionisation energy on the molecular geometry, [5,6] and the combs therefore mark the energy regions that correlate with the singlet and triplet excited states of the bound $CS_2$ molecule.

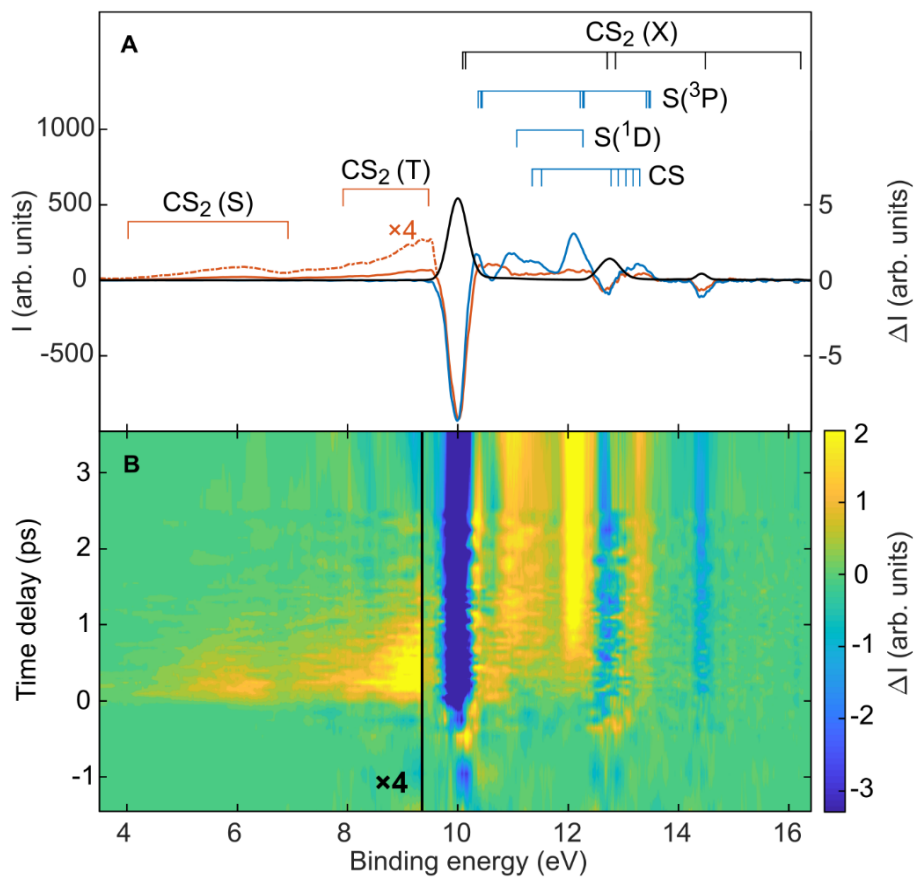

**Figure 2.** Pump-probe photoelectron spectra. (A). Ground state photoelectron spectrum (black line, left axis). Average background-subtracted spectra obtained at early (40 – 540 fs, red line, right axis) and late (4.04 – 10.54 ps, blue line, right axis) times following a 6.2 eV pump. Negative features correspond to ground state bleaching while positive features correspond to new bands in the photoelectron spectrum. The combs mark assignments based on known ionisation limits of the produced fragments. (B) False colour surface map showing the changes in the background-subtracted photoelectron spectrum as a function of pump-probe time delay.

The time-dependent changes in the background-subtracted photoelectron spectrum are also plotted in Figure 2B. Excitation leads to a depletion of the bands associated with the ground state $CS_2$ molecules as well as a commensurate increase in signal associated with population of the excited singlet state. Ionisation of the excited singlet state into the ground ion state of the molecule produces a photoelectron band starting at 4 eV. The photoelectron band stretches between 4–7 eV due to large changes in the ionisation energy associated with geometry changes in $CS_2$. [5,6] The damping of the vibrational motion leads to the narrowing of the band towards higher binding energies over the few hundred fs in which significant population remains within the excited singlet states. As the singlet state population decays we observe a new band at binding energies 7–10 eV. We preliminarily assign this signal to population in the triplet state following intersystem crossing. The band maximum is seen to shift to higher binding energies as the molecule dissociates. The shifting of the photoelectron bands to the asymptotic dissociation limits associated with the final $CS(X\ ^1\Sigma^+)$, $S(^3P)$ and $S(^1D)$ dissociation products creates transient intensity over a large energy range. The strongest product signal is observed between 12.1–12.4 eV and contains contributions from both the $S(^1D)$ and $S(^3P)$ dissociation products.

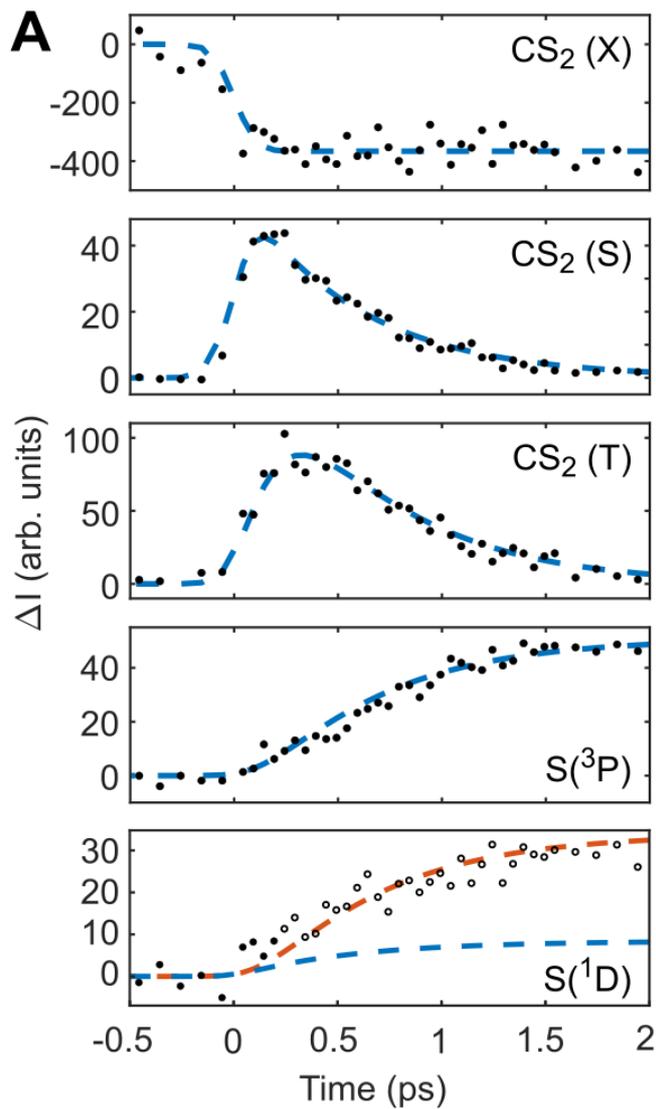
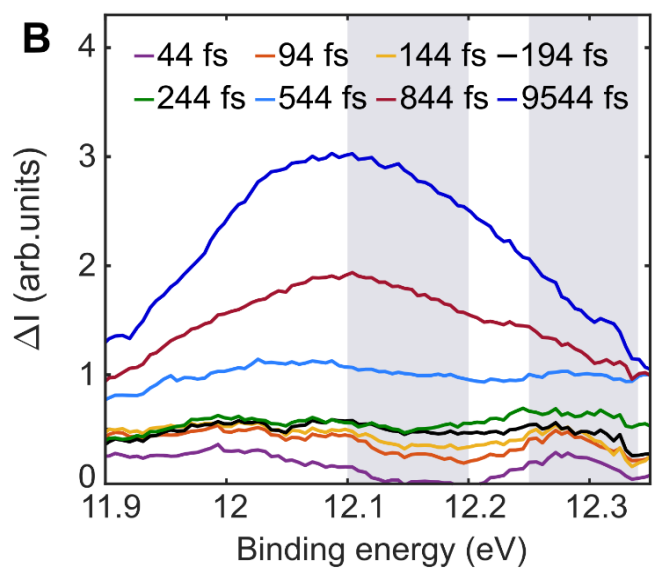

**Figure 3. Pump-probe kinetics.** (A) Time-dependent photoelectron intensity plots corresponding to: the ground state, $CS_2(\tilde{X})$; the excited singlet states, $CS_2(S)$; the excited triplet states $CS_2(T)$; the singlet dissociation products, $S(^1D)$; and the triplet dissociation products, $S(^3P)$. For $S(^1D)$ solid points are used where the data is reliable and hollow symbols once the $S(^3P)$ state contributes to the measured intensity. Dashed blue lines are the results of fits to the kinetic model. The dashed orange line in the $S(^1D)$ is the kinetic fit associated with triplet state formation and confirms the origin of the delayed rise in this region. (B) An expanded view of the photoelectron spectrum in the region covered by the $S(^1D)$ and $S(^3P)$ dissociation products. Each spectrum is the average of three time delays with the average delay provided in the legend. The highlighted regions cover those used for the $S(^1D)$ and $S(^3P)$ plots in (A).

To highlight the changes in the photoelectron spectrum, we plot the integrated electron count over energy regions that correlate with specific states in the dynamic process in Figure 3A. The selected regions cover the ground state ($CS_2(\tilde{X})$, 9.8–10.1 eV), singlet excited states ($CS_2(S)$, 4.0–6.9 eV), triplet excited states ($CS_2(T)$, 7.9–9.4 eV), triplet dissociation product ($S(^3P)$, 12.1–12.2 eV), and singlet dissociation product ($S(^1D)$, 12.25–12.33) eV. Analysis of the features associated with dissociated population is complicated by the density of the spectrum and by features associated with transient structures. We plot an expanded view of the photoelectron spectrum over the 11.90–12.35 eV region, associated with the atomic S signal, used in Figure 3A at various times after excitation in Figure 3B. Based on known ionisation limits of the S atom, in this energy region we expect signals due to the $S^+(^2P) \leftarrow S(^1D)$ transition at 12.26 eV, and $S^+(^2D) \leftarrow S(^3P)$ transitions between 12.1-12.2 eV. At early times (< 200 fs) we see a rapid increase in the $S(^1D)$ signal at 12.26 eV that is the major feature in the spectrum. There is also significant intensity at energies below 12.1 eV, which shifts to higher energies with time. At increasing delay-times, the maxima of the lower energy feature shifts to energies associated with the $S(^3P)$ product that eventually dominates the spectrum including the region initially dominated by the $S(^1D)$ signal.

A measure of the $S(^3P)$ and $S(^1D)$ yields can be obtained by integrating over the energy windows highlighted in Figure 3B. The trace obtained for the $S(^1D)$ is dominated by signal associated with the $S(^3P)$ product at long times and is therefore only reliable for the first 200-250 fs following excitation while the $S(^3P)$ yield remains small. The integrated electron count over the highlighted regions are plotted in Figure 3A. Reliable data points for the $S(^1D)$ channel are plotted as solid symbols and those that have a significant contribution from the $S(^3P)$ as hollow. The plots therefore provide a quantitative measure of the $S(^3P)$ yield and a qualitative measure of the time dependence of the production of $S(^1D)$ fragments. Based on the changes in the spectrum and in the integrated yields, it appears that $S(^1D)$ is produced more rapidly than $S(^3P)$, but with lower yield. After the initial formation (at times < 100 fs), we observe no measureable increase in signal until after ~500 fs. This increase appears has the same time profile as the triplet state products (shown by the dashed orange line in figure 3A) and is due to the overlapping signal associated with the triplet dissociation product. Such that despite significant population remaining in the excited singlet states for several hundred femtosecond after the initial rise, we see no further singlet dissociation products formed.

The dynamic traces are fitted to a kinetic model (see SI [24]) that provides effective lifetimes of the excited states. The results are overlaid on the experimental data in Figure 3A as blue dashed lines. The fits provide a singlet excited state lifetimes of 570 fs, which correlates with the rising triplet state population suggesting the excited singlet state predominantly undergoes ISC. The triplet state then has a lifetime of 170 fs that correlates with the appearance of the $S(^3P)$ dissociation products. The delayed appearance of the features between 7.9 – 9.4 eV and their correlation with the formation of the atomic triplet state product confirming our earlier assignment. The dynamics of the singlet state are somewhat more complex and do not fit a first order kinetic model. The significant intensity in the singlet dissociation products at very early times (>100 fs) does not correlate with the singlet or triplet state populations or effective lifetimes. The initial step correlates with the timescale of the $CS_2$ bending motion, suggesting an initial impulsive dissociation process. The singlet dissociation yield then stays roughly constant suggesting the singlet population is stabilised, most likely due to the rapid IC dynamics with the manifold of accessible singlet states. The later time increase in dissociation yield seen in the singlet channel region matches that associated with triplet state formation as shown by the orange dashed line in Figure 3A.

Based on experimental data and the calculations we suggest the following mechanism. Upon excitation, dissociation along the singlet channel competes with rapid IC, which redistributes the population among the singlet states. Note in Fig. 1 how bending of the molecule brings the electronic states into close vicinity, with additional curve crossings to complement those in the radial coordinate (See SI [24]). The IC process is associated with transfer to the $4^1A'$ electronic state driven by rapid bending and stretching of the molecule. The $4^1A'$ state has a high dissociation limit and acts as a storage mode for the molecule reducing the likelihood of further dissociation. The changes in electronic and geometric structure result in a dynamic stabilisation that reduces the likelihood of further singlet dissociation, strongly diminishing the singlet dissociation process within 200 fs. In the meantime, spin-orbit coupling leads to ISC and population transfer to the triplet states, giving an effective singlet state lifetime of 560 fs. Based on the large energy shift associated with the ISC process and the calculated potential energy curves, we suggest that the triplet states populated are $1^3A'$, $1^3A''$ and $3^3A''$. These have low barriers to dissociation across a wide range of molecular geometries leading to the formation of CS + S($^3$P) triplet product, with the bound triplet state lifetime of approximately 180 fs. The branching ratio between singlet and triplet products is therefore controlled by the strength of the spin-orbit coupling, *and*, also by the extremely fast non-adiabatic dynamics in the excited singlet states.

Time-resolved XUV-PES provides a detailed view of chemical processes and highlights the complex interplay between electronic and nuclear dynamics even in structurally simple molecules, *and*, the importance of measuring the entire reaction pathway when defining reaction mechanisms. Providing a similarly global experimental and theoretical measure of the dynamics in larger systems, while challenging, is key to increasing our understanding of far-from-equilibrium photochemical dynamics. Theoretical comparisons will require the calculation of the full dynamical process and how these dynamics project onto the photoioinisation measurement. Experimental developments in few-cycle, high repetition-rate laser systems, will allow us to combine XUV-PES with coincidence detection that has the potential to measure dynamics with molecule specific spectra obtained with attosecond temporal resolution. The opportunities provided by the combination of experimental techniques based around time resolved XUV PES, X-ray/electron scattering [29–31] and element-specific core-level spectroscopies, [20,21,32] and associated theory, brings with it a new age in chemical understanding, where mechanisms are assigned based on direct interrogation of the full reaction pathway.

**Acknowledgments:**

The authors thank the STFC for access to the Artemis facility and funding from the European Commissions' Seventh Framework Programme (LASERLAB-EUROPE, grant agreement 228334). AK and RSM acknowledge the Leverhulme Trust (RPG-2013-365) for research support and for studentship (ADS) and postdoc (MT) funding. RSM thanks the Royal Society for funding (UF100047, UF150655 and RG110310), and AK thanks the European Union (FP7-PEOPLE-2013-CIG-NEWLIGHT). DB and EMW thank the Universities of Edinburgh and Southampton for studentships. DAH has been supported by the excellence cluster "*The Hamburg Center for Ultrafast Imaging-Structure, Dynamics and Control of Matter at the Atomic Scale*" of the Deutsche Forschungsgemeinschaft (CUI, DFG-EXC1074), by the European Research Council through the Consolidator Grant Küpper-614507-COMOTION, and by the Helmholtz Gemeinschaft through the "*Impuls- und Vernetzungsfond*". The computational work reported used the ARCHER UK National Supercomputing Service (http://www.archer.ac.uk) and the Edinburgh Compute and Data Facility (ECDF) (http://www.ecdf.ed.ac.uk). We thank Andrés Moreno Carrascosa (Edinburgh) and Phil Rice (Artemis) for technical assistance. Data supporting this study are openly available from the University of Southampton repository at http://doi.org/10.5258/SOTON/D0455.